 \documentclass[final,1p,times,twocolumn]{elsarticle}

\usepackage{amssymb}
\usepackage{amsmath}

\journal{Nuclear Physics B}

\begin{document}

\begin{frontmatter}



\title{Mathematical and Computational Modeling of Amoeboid Cell Crawling}


\author{Sergio Alonso} 

\affiliation{organization={Department of Physics, Universitat Politècnica de Catalunya—BarcelonaTech},
            addressline={Carrer de Jordi Girona 1-3}, 
            city={Barcelona},
            postcode={08034}, 
            country={Spain}}

\affiliation{organization={Institute for Research and Innovation in Health (IRIS), Universitat Politècnica de Catalunya—BarcelonaTech},
            addressline={Av. Diagonal, 647}, 
            city={Barcelona},
            postcode={08028}, 
            country={Spain}}

\author{Carsten Beta} 

\affiliation{organization={Institut für Physik und Astronomie, Universität Potsdam},
            addressline={Karl-Liebknecht-Straße 24/25}, 
            city={Potsdam},
            postcode={14476}, 
            country={Germany}}

\affiliation{organization={Nano Life Science Institute (WPI-NanoLSI), Kanazawa University},
            addressline={Kakuma-machi},
            city={Kanazawa 920-1192},
            country={Japan}}

\begin{abstract}
Amoeboid motion is a dynamic mode of cell motility essential for processes such as the immune response and wound healing. This review examines recent developments in the mathematical and computational modeling of amoeboid crawling, focusing on the interplay between intracellular biochemical signaling and the physical mechanics of the cell membrane. We discuss the core components of cell motility and the integration of chemical and mechanical guidance cues suchg as chemotaxis and curvotaxis. We evaluate a range of modeling frameworks, from simple stochastic descriptions of center of mass motion to more complicated phase-field, finite-element methods and Potts models that capture complex cell shape deformations. Finally, we highlight emerging challenges, such as modeling interactions with complex topographies and large-scale multicellular coordination, as important steps toward a better understanding of cell locomotion.
\end{abstract}






\end{frontmatter}



\section{Introduction}
\label{sec1}

Living cells are inherently motile, navigate their surroundings to evade predators, seek nutrients, or seek more favorable environments. Although bacteria typically navigate using temporal sensing to detect chemical gradients over time, much larger eukaryotic cells possess the structural complexity to sense and respond directly to spatial gradients across their bodies.

\begin{figure}[t]
\centering
\includegraphics[width=0.70\textwidth]{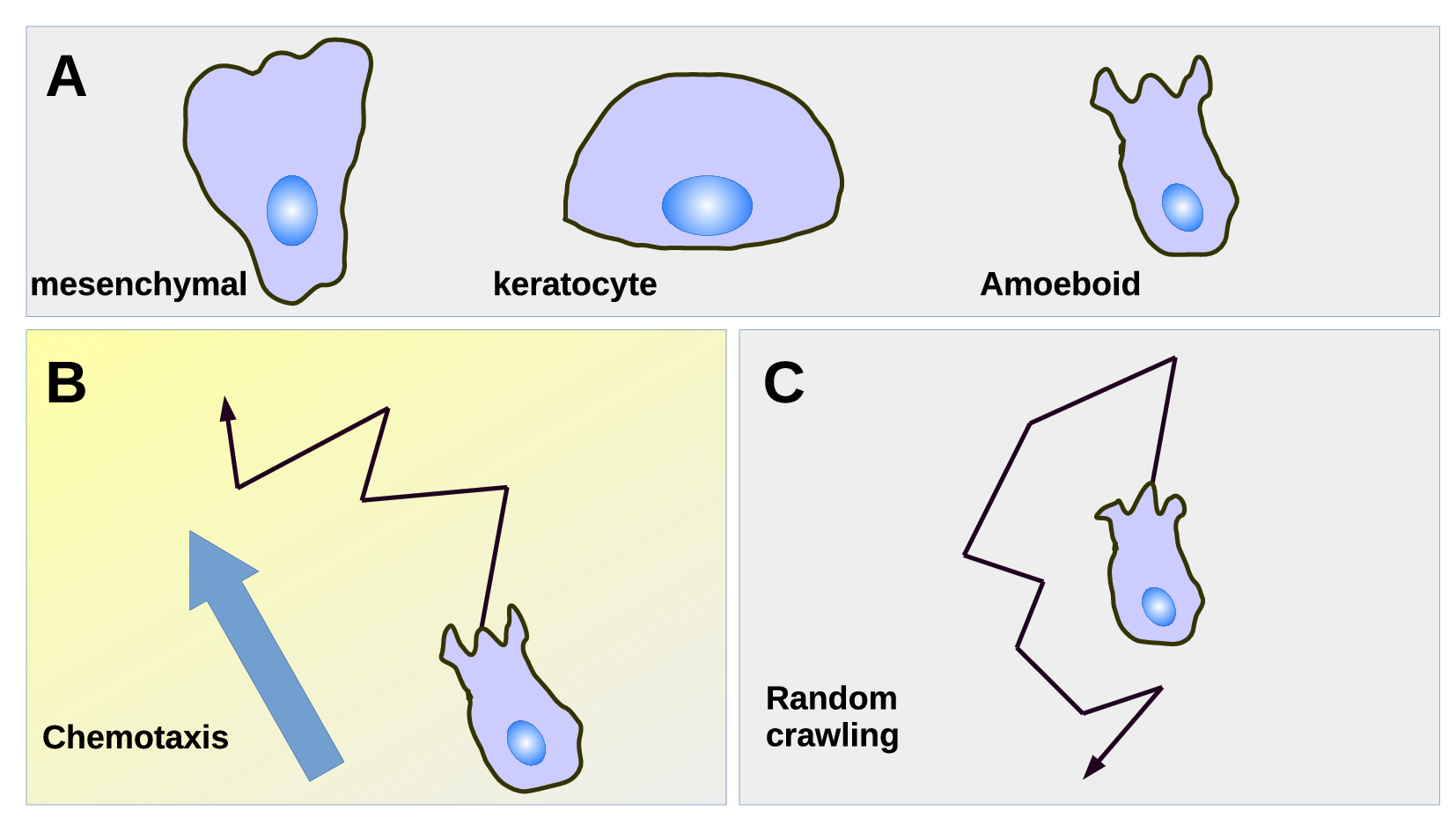}
\caption{Types of cells and motion strategies (A): slow  and large mesenchymal cells ($1 \mu m /min$, $50 \mu m$), fast and large keratocytes  ($30 \mu m /min$, $40 \mu m$), and fast and small amoeboid cells  ($15 \mu m /min$, $20 \mu m$). Chemotactic (B) and random (C) motions of amoeboid cells.}\label{fig1}
\end{figure}

On two-dimensional substrates, migratory phenotypes vary significantly between cell types, see Fig.\ref{fig1}. Some cells exhibit mesenchymal migration, characterized by relatively low velocities and strong substrate attachment. In contrast, others, such as fish epithelial keratocytes, adopt stable fan-like morphologies that remain essentially stationary in shape as they glide. Furthermore, migratory fibroblasts, neutrophils, and the social amoeba {\it Dictyostelium discoideum} exhibit highly dynamic amoeboid migration \cite{bodor2020cell,toscano2024methods}. In these systems, the ability of cells to adhere to a surface and the resulting efficiency of its propulsion are fundamentally determined by the precise coordination between intracellular contractile forces and extracellular traction forces \cite{mogilner2023crawling}.

Cell locomotion is fundamentally governed by the biochemistry of the actin cytoskeleton \cite{banerjee2025signaling} and can occur independently of immediate genetic regulation; indeed, even enucleated cells (cells from which the nucleus has been removed) exhibit near-normal migratory behavior \cite{de2025migratory}. Consequently, motility is primarily determined by the spatial distribution, function, and turnover of actin filaments, which dynamically move, resulting in a rich phenomenology on actin waves \cite{beta2023actin}. 
These proteins polymerize to exert physical pressure against the cell membrane, driving the protrusive forces necessary for movement \cite{yochelis2022versatile,kay2024making}.
A finite pool of actin polymerizes against the cell membrane at multiple locations, generating protrusions known as pseudopodia. These structures compete as they extend in the direction of local force until a single dominant pseudopod emerges, typically at the front of the cell, to establish the primary axis of motion \cite{alonso2025persistent}.

Giant cells can be generated from populations of individual cells through electrofusion, which consists of the application of an electric pulse. These cells serve as an ideal experimental system for investigating internal wave dynamics; by significantly expanding the cytoplasmic volume, the influence of boundary constraints and compartmentalization is minimized \cite{yang2022cortical}. 

Individual cells move and coordinate their dynamics with the collective to generate complex behaviors and emergent tissue-level dynamics \cite{buttenschon2020bridging}. A paradigmatic example of this cooperation is {\it Dictyostelium discoideum}. 
In this system, single-cell motion, driven by excitable self-organized biochemical networks, is triggered by the secretion of cAMP. This signaling initiates an oscillatory self-organization and subsequent aggregation, culminating in morphogenesis and coordinated movement of the multicellular slug \cite{schaap2021environmental}.

The paper is organized as follows, we begin with a concise description of the biochemistry and biophysics of cell crawling, starting with the mechanisms of cell polarization that trigger the locomotion process. We then review the various mathematical methods, models, and computational tools used to characterize cell crawling, with a focus on intercellular interactions and physical constraints. Finally, we outline future perspectives and emerging challenges in the modeling of cell locomotion.












\section{Cells in motion}
\label{sec2}

\paragraph{Actin-driven motility}
Actin-based locomotion is one of the most widespread forms of cell motility in the living world~\cite{alonso2025physical}.
It primarily relies on actin polymerization, where the growth of actin filaments generates forward-pushing forces to create membrane protrusions, such as pseudopodia and lamellipodia.
Actin polymerization is orchestrated by regulatory proteins, such as the small GTPases Rac, Rho, and Cdc42, that activate nucleation-promoting factors, including formins and the Arp2/3 complex via WASP-WAVE family proteins.
At the same time, depolymerization promoting factors, such as ADF/cofilin, ensure disassembly and turnover of actin monomers to maintain continuous treadmilling of actin filaments.
For more details of the underlying molecular mechanisms, see classical reviews on actin dynamics in cell migration, such as~\cite{blanchoin2014actin,schaks2019actin} and references therein.
For stable directional migration, cell polarity is required, where the actin-driven formation of membrane protrusions is spatially confined to the leading edge of the cell~\cite{bera2025cell}.
On the opposite side, myosin motors bind to the actin filaments and generate contractile tension to retract the rear part of the cell.
This asymmetric organization of cellular components establishes a polarized front-rear axis
and is maintained through mutual inhibition of signaling pathways, polarized vesicle transport, and membrane tension gradients~\cite{bera2025cell,de2024follow}.
Finally, adhesion complexes that anchor the actin cytoskeleton to solid substrates or to the extracellular matrix transmit the traction forces ultimately required for locomotion~\cite{alonso2025physical}.

\paragraph{Eukaryotic chemotaxis}
In many fundamental biological processes, such as wound healing, immune responses, and embryonic development, cell migration is guided by chemical cues.
The directed migration of cells along chemical gradients, commonly referred to as chemotaxis, integrates random motility, polarity formation, and directional sensing, and has been intensively studied in neutrophils and model organisms such as \textit{Dictyostelium discoideum}.
The binding of chemoattractant molecules to G-protein coupled receptors (GPCRs) activate heterotrimeric G proteins that trigger a network of parallel downstream signaling pathways at the leading edge.
These include the activation of Ras GTPases, PI3K-mediated PIP3 production, and TorC2-PKB signaling, all of which coordinate the activity of the actin machinery that is required to drive the formation of membrane protrusions.
Simultaneously, cGMP signaling activates myosin II to generate the contractile forces necessary for membrane retraction.
Together, this complex signaling network forms a nonlinear dynamical system that incorporates adaptation, amplification, and feedback mechanisms.
It sensitively detects small differences in receptor occupancy across the cell membrane (spatial sensing) and amplifies shallow chemoattractant gradients, over a wide range of ambient background concentrations, into a strong intracellular asymmetry.
In this polarized configuration, protrusive actin polymerization is confined to the cell front pointing towards higher chemoattractant concentrations, while contractile actomyosin activity retracts the back of the cell.
Classical review papers, such as~\cite{van2004chemotaxis,swaney2010eukaryotic,devreotes2017excitable}, provide detailed overviews of chemotactic signaling pathways in eukaryotic cells.
For selected aspects, such as Ras and GPCR signaling, more recent reviews are also available, see e.g.~\cite{kamimura2021dynamical,xu2022ras}.

\paragraph{Other forms of taxis}
Besides gradients of soluble chemoattractants, cells may also navigate in other directional cues, such as substate-bound chemical gradients (haptotaxis), electric fields (galvanotaxis), or gradients in substrate stiffness (durotaxis).
Also the geometrical properties of the substrate may guide cell migration, in particular, surface topography (topotaxis) or substrate curvature (curvotaxis), see~\cite{shellard2020all,sengupta2021principles} and references therein.
In most cases, the different forms of environmental stimuli converge on a common set of downstream effectors and conserved signaling modules, involving small GTPases, such as Rac, Rho, and Cdc42, and the actin cytoskeleton.
Together, they establish cellular polarity along the axis of the external stimulus, with a clearly defined leading edge on one side, where typical cell front markers accumulate and membrane protrusions are formed, and a retracting back on the other.
In the case of multiple conflicting external signals, cells integrate information from different input channels but may prioritize one of them~\cite{bull2022actin}.




\begin{figure}[t]
\centering
\includegraphics[width=1.0\textwidth]{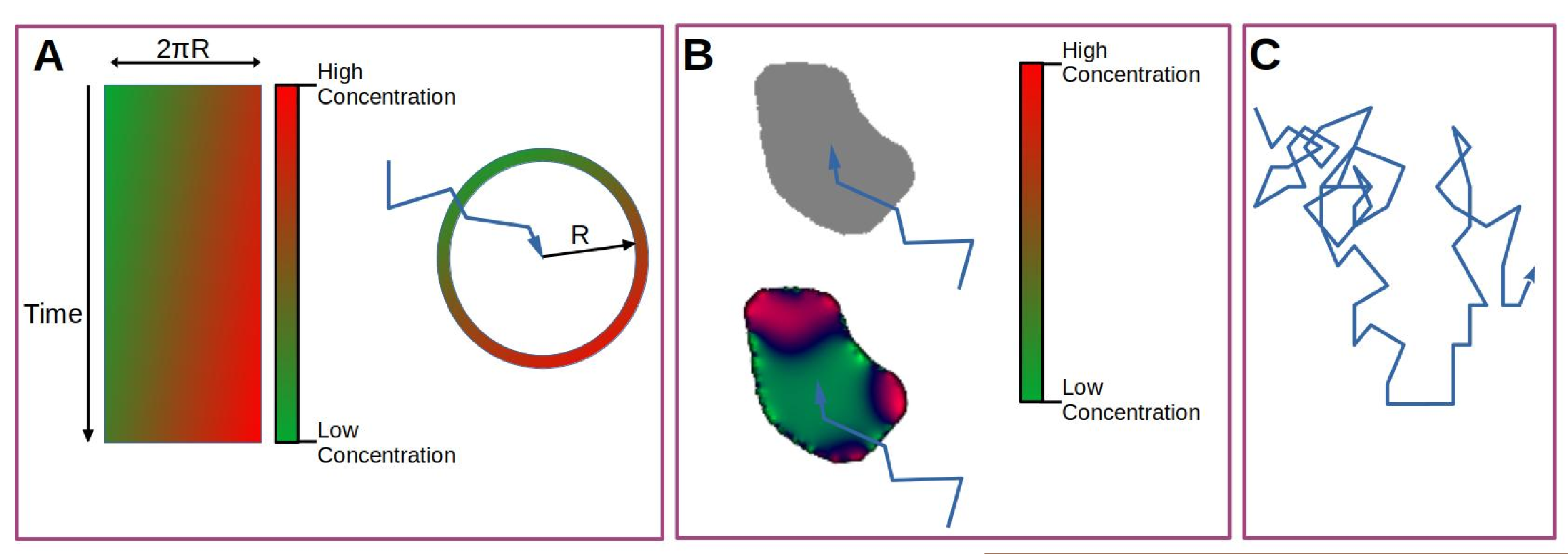}
\caption{Modeling approaches to cell motion: (A) evolution of one dimensional pattern formation at the cell membrane, see Section \ref{subsec3.1}, (B) two dimensional pattern formation including shape deformation of the borders and the internal pattern formation, see Section \ref{subsec3.2}, (C) center of mass, see Section \ref{subsec3.3}.} \label{fig3}
\end{figure}

\section{Models of cell crawling}
\label{sec3}

As briefly outlined in the previous section, the crawling of eukaryotic cells is a complex multiscale process.
Molecular processes at the subcellular level, including receptor-mediated signaling pathways, actin polymerization dynamics, and cytoskeletal motor activity, drive coordinated deformations of the cell contour that ultimately result in a translocation of the cell's center of mass.
Out of these multiple layers of complexity (intracellular dynamics, morphological changes in the cell contour, and center-of-mass locomotion), mathematical models of cell crawling often focus on one of them, while including the others only in a coarse-grained fashion or even neglecting them, depending on the questions they are targeting.
Here, we will briefly introduce the different modeling approaches, see Fig.\ref{fig3}, emphasizing recent developments in this field.

\subsection{Modeling pattern formation at the membrane}
\label{subsec3.1}

Living cells are typically observed by optical microscopy, revealing various structural features depending on the position of the focal plane within the cell.
In motile cells, intracellular dynamics is visualized mainly by introducing fluorescent variants of proteins that are involved in actin cytoskeletal dynamics and/or the respective upstream signaling pathways.
The space-time dynamics of these components is mostly confined to the cell membrane and the adjacent cortical zone, where it can be observed by laser-scanning confocal microscopy, either during random cell migration or in the presence of extracellular cues.
Because these proteins interact through biochemical reactions and are transported primarily by diffusion, partial differential equations (PDEs) of reaction-diffusion type are the most common mathematical framework to model the intracellular protein dynamics in motile cells.

\paragraph{Pattern formation at the cell border}
In many cases, experimental data is recorded in a focal plane that is placed in the center of the cell body, capturing the cytosolic region surrounded by the cell membrane.
In the focal plane, the membrane appears as a one-dimensional closed contour that evolves dynamically over time as the cell deforms. 
The membrane localization of signaling molecules and the corresponding activity of the actin cytoskeleton that drive the contour dynamics, can be experimentally traced along this one-dimensional domain.
To simulate the space-time dynamics of these concentration patterns, a reaction-diffusion system can be implemented on a one-dimensional domain with periodic boundary conditions that captures various mechanisms of pattern formation.
%
A typical set of reaction-diffusion equations reads
\begin{equation}
 \frac{\partial\boldsymbol{c}}{\partial t} = \boldsymbol{F}(\boldsymbol{c}; \boldsymbol{p}) + \boldsymbol{D} \nabla^2 \boldsymbol{c}
  \label{eq:rd}
\end{equation}
where $\boldsymbol{c}=(c_1,c_2,...,c_n)$ are the concentrations of interacting molecules, $\boldsymbol{p}=(p_1,p_2,...,p_m)$ the parameters of the reactions $\boldsymbol{F}$ that are in general nonlinear, and $\boldsymbol{D}$ is the diffusion matrix, which typically takes diagonal form.
In the case of a simple activator-inhibitor system, Eqs.~\eqref{eq:rd} reduces to two coupled equations for $c_1$ and $c_2$.

Numerical integration of Eqs.~\eqref{eq:rd} yields characteristic spatiotemporal patterns, such as traveling waves, standing waves, and Turing-like structures that can be compared to fluorescence microscopy recordings of biochemical components, such as PIP$_3$ or Ras~\cite{bhattacharya2020traveling}. 
In particular, studies relying on cells treated with Latrunculin~A, a drug that promotes actin depolymerization, have a long history.
Due to the absence of an intact actin cytoskeleton, cells treated with Latrunculin~A maintain a constant spherical shape allowing for a direct comparison of signaling patterns along the circular membrane contour to a one-dimensional reaction-diffusion system.
Here, recent work has focused on the interplay between an excitable Ras signaling system, bistable phase separation of PIP$_2$ and PIP$_3$, and stochastic fluctuations that can trigger excitable and oscillatory waves at the cell membrane, for a recent review, see~\cite{matsuoka2024spontaneous}.

\paragraph{Influence of mass conservation}
The pattern forming biochemical reactions take place in the confined volume of a cell.
Because many of these reactions occur much faster than protein biosynthesis, the total quantity of some of the biochemical components remains constant over the characteristic timescale of the observed phenomena.
Such mass conservation constraints fundamentally influence the pattern formation process and have been a major focus of recent research in this field~\cite{weyer2025protein}.
For example, mass conservation can lead to global constraints arising through the cytosol, which can be incorporated via integral terms into the reaction-diffusion system~\eqref{eq:rd}, 
\begin{equation}
c_{i}^{cyt} = c_{i}^{tot} - \int c_{i} dx 
     \label{eq3.2b}
\end{equation}
where $c_{i}^{tot}$ is the constant total concentration of component $c_i$ and $c_{i}^{cyt}$ corresponds to its cytosolic concentration that may enter the kinetic functions $\boldsymbol{F}$ as an additional time-varying parameter.
Such constraints are regularly used in recent models of cell crawling~\cite{imoto2021comparative}.
In particular, they have been implemented in a model of Ras activation to investigate the role of stochastic terms versus deterministic bistable dynamics~\cite{moreno2022mass}, showing that a mass-conservation constraint can enhance the robustness of the underlying polarization mechanisms.

Mass conservation can be also taken into account by balancing some of the kinetic terms in the nonlinear functions $\boldsymbol{F}$, so that the overall number of the respective molecular species is conserved, see for example
\begin{eqnarray}
 \partial_t c_1 \,=& &f(\boldsymbol{c}; \boldsymbol{p}) + D_1 \nabla^2 c_1\nonumber \\
 \partial_t c_2 \,=&-\hspace{-4mm}&f(\boldsymbol{c}; \boldsymbol{p}) + D_2 \nabla^2 c_2 \label{eq:rd_conserved} \\
 \partial_t c_3 \,=& &g(\boldsymbol{c}; \boldsymbol{q}) + D_3 \nabla^2 c_3\, ,\nonumber
\end{eqnarray}
with $\boldsymbol{c}=(c_1,c_2,c_3)$, where $c_1+c_2=c_{tot}$ is constant over time.
Models of the form~\eqref{eq:rd_conserved} have been introduced to study pattern formation in the actin cytoskeleton, see~\cite{beta2023actin} for a recent review.
Here, the mass-conserved part is motivated by upstream small GTPase signaling, where active and inactive forms of the GTPase are converted into each other.
The mass conservation constraint introduces distinct dynamical features that are generally not present in arbitrary forms of model~\eqref{eq:rd}.
For example, bifurcation analysis of a mass conserved activator-inhibitor system of type~\eqref{eq:rd_conserved} has revealed a regime, where traveling waves and excitable pulses coexist, similar to experimental observations in oversized \textit{Dictyostelium} cells~\cite{yochelis2022versatile}.
Also the bistable switching between polarized (fan-shaped) and disordered states of \textit{Dictyostelium} migration could be explained with a similar model~\cite{hughes2025dissipative}.

A more complex variant of a mass-conserving reaction-diffusion system successfully captured rotation, polarization, and oscillations of Rac-enriched domains observed in fluorescence microscopy recordings of \textit{Dictyostelium} cells~\cite{vsovstar2024oscillatory}. 
%
Note also that the concept of Absolute Concentration Robustness (ACR) is related to the conservation of molecular species~\cite{shinar2010structural}.
Introducing an ACR mechanism into the signaling system of motile cells was proposed as possible mechanisms to reduce concentration fluctuations and thus the likelihood of threshold crossings in excitable signal transduction networks~\cite{biswas2023enhanced}.


\paragraph{Pattern formation at the basal membrane}
%
The reaction-diffusion equations~\eqref{eq:rd} can be extended to two-dimensional domains, representing the inner face of the plasma membrane.
In most cases, two-dimensional simulations are carried out for comparison with experimental recordings of protein dynamics at the basal substrate-attached membrane, which is readily accessible by fluorescence microscopy due to its flat geometry. 
A common focus of many established reaction-diffusion models that were explored on a two-dimensional domain is the study of cortical wave patterns that are often related to the formation of membrane protrusions and to motility~\cite{beta2023actin}.


More recent work focused on dynamic partitioning that results in large-scale compartmentalization of lipid-anchored and integral membrane proteins during membrane polarization~\cite{banerjee2023dynamic}.
Also, additional feedback loops were integrated to account for mechanical effects of the cytoskeleton that influence signal transduction and polarity formation~\cite{kuhn2025complementary}.
Specifically, two types of negative feedback were considered, local inhibition near the site of activity and global inhibition affecting the entire cell.
Although both types of feedback can stabilize the cell front, global feedback was more efficient in preventing the formation of multiple leading edges~\cite{banerjee2025spatial}.
Besides conventional reaction-diffusion models, the reaction-diffusion master equations have been used to include the noise that arises from the stochastic descriptions of the reaction and diffusion terms~\cite{biswas2021three}.
Here, G-protein coupled receptor signaling (GPCR) and a local excitation-global inhibition mechanism (LEGI), together with a signal transduction excitable network (STEN), have been implemented in a static hemispherical domain to mimic a cell in which the actin cytoskeleton was disrupted by treatment with Latrunculin~A.




\subsection{Modeling cell shape dynamics}
\label{subsec3.2}

Space-time patterns in signaling activity, together with the downstream cytoskeletal machinery, drive the shape dynamics of motile cells.
The modeling of the evolution of the cell shape thus requires a robust method that describes the dynamics of the cell boundary and couples it to the intracellular processes.
For this, several computational approaches have been proposed in the past~\cite{dinapoli2021tools}.
In most cases, they focus on a two-dimensional representation of the cell, where dynamical patterns on the basal membrane drive deformations of the surrounding boundary.
Here, we focus on recent developments regarding two of the most prominent strategies: the use of an additional phase field to define the cell shape and mechanically inspired interface models relying mostly on the finite element method (FEM) to solve the dynamics directly on the cell surface.
Other alternative methodologies, such as the Cellular Potts models, are also briefly discussed.
For a comparison of the different methods, see Table~\ref{Tab1}.

\paragraph{Phase field model}
%
One of the most widely used methods for modeling cell shape deformations driven by internal biochemical dynamics is the phase-field approach~\cite{moure2021phase,stinner2020mathematical}.
In this framework, an auxiliary field is defined, taking a value of 1 inside the cell and 0 outside.
The field is defined to enforce no-flux boundary conditions as the cell perimeter evolves.
In the sharp-interface limit, this method precisely tracks the boundary between the interior and the exterior environment, see Table~\ref{Tab1}.
The equation that governs its dynamics typically incorporates terms that represent surface tension, area conservation, and active stresses that represent the mechanical activity of the actin cytoskeleton.
It has been widely applied to different forms of cell migration~\cite{aranson2016physical}.

\begin{figure}[t]
\centering
\includegraphics[width=0.80\textwidth]{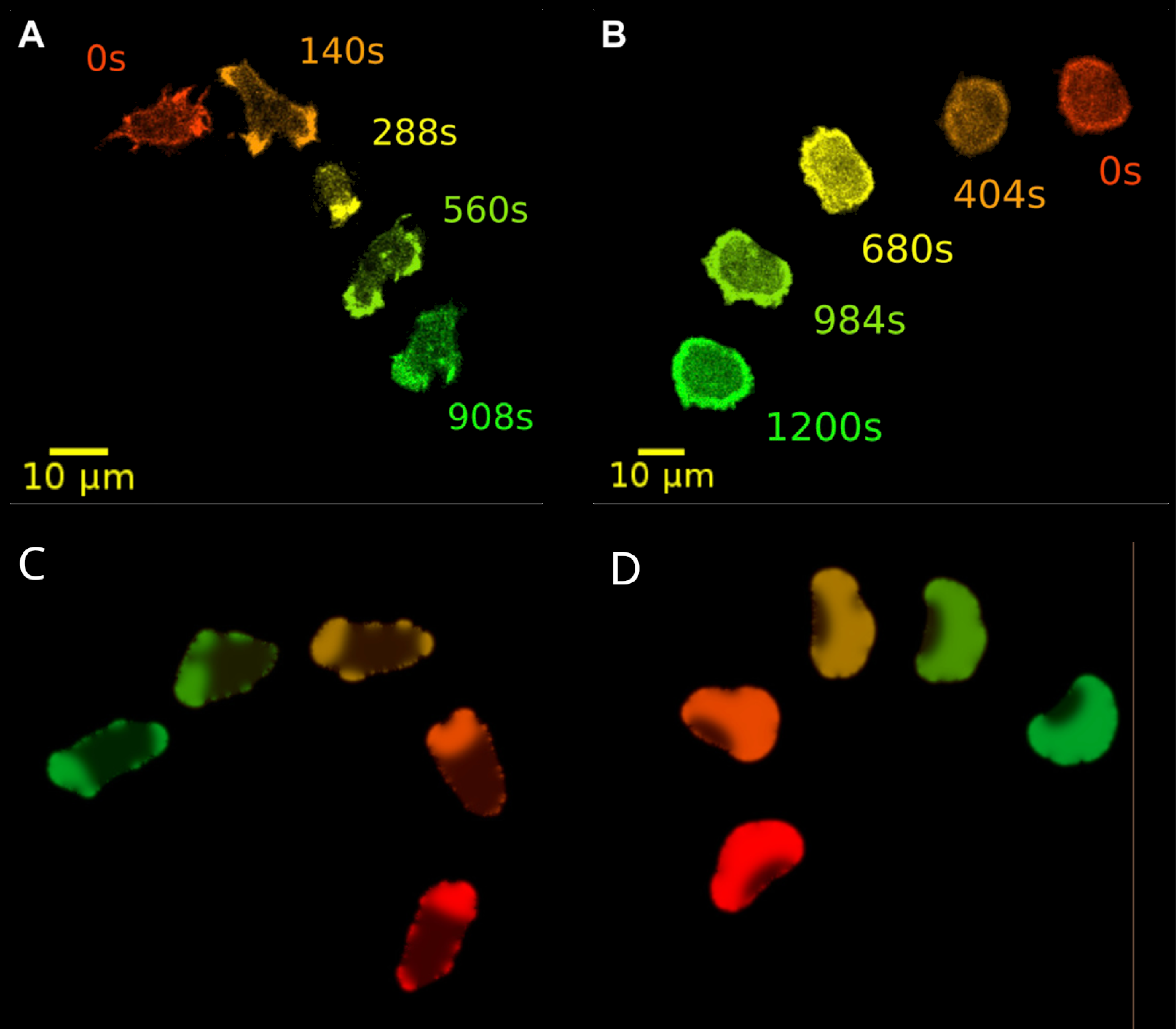}
\caption{Types of cell motion strategies for {\it Dictyostelium discoideum}: Amoeboid (A) and fan-type (B) motions, where color corresponds with the expression of Lifeact-GFP as a marker for filamentous actin; and for numerical simulations with phase field: Amoeboid (C) and fan-type (D) motions, where color corresponds with the internal field from a stochastic reaction-diffusion process. Time is color coded from red, orange and
yellow to green. }\label{fig2}
\end{figure}

Recent modeling studies, which rely on phase field approaches, focused on migratory plasticity.
Specifically, the coupling between intracellular biochemical reaction-diffusion patterns and active stresses that deform the phase field boundary can account for various modes of motility in {\it Dictyostelium discoideum}, ranging from typical amoeboid movement to fan-shaped and oscillatory dynamics, depending on the total abundance of specific biochemical components, see Fig.~\ref{fig2}~\cite{alonso2018modeling,cao2019plasticity,moreno2020modeling}. 
Spontaneous transitions between these distinct migratory modes have been observed experimentally and were successfully reproduced using models that incorporate bistable dynamics~\cite{moldenhawer2022spontaneous,hughes2025dissipative}.
Similar amoeboid, fan-shaped, and oscillatory patterns have been obtained by coupling a phase field with random protrusion distributions and solving the Stokes equations for internal fluid dynamics \cite{ghabache2021coupling}.
These various modes of motility can be further analyzed using Principal Component Analysis to quantitatively characterize the transitions between them~\cite{imoto2021comparative}, a method validated against the diverse dynamics of {\it Dictyostelium discoideum}, HL60 cells, and keratocytes.
Furthermore, it is possible to restrict the dynamics of specific molecular components to the cell membrane by defining the boundary through the gradient of the phase field~\cite{zhang2025phase}.
This approach recovers fundamental amoeboid movement, as well as the "bipedal" dynamics frequently observed in {\it Dictyostelium discoideum}~\cite{van2021short}.


\paragraph{Mechanical interface descriptions}
%
The finite element method (FEM) is ideally suited to model mechanical deformations driven by elastic or viscous contributions~\cite{stinner2020mathematical}. 
This framework allows for simultaneous numerical integration of reaction-diffusion equations of biochemical components alongside the advection and viscoelastic properties of the membrane, see Table~\ref{Tab1}. 
It triangulates the moving interface to track the dynamics of the cell boundary.
It can also be combined with the previously described approach to solve phase field models with FEM~\cite{moure2021phase}.
This approach has been successfully applied to model the Rac dynamics coupled with the viscoelastic characteristics of neutrophils~\cite{zmurchok2020membrane}.
Models of signaling patterns have also been combined with intracellular fluid dynamics of the cytoplasm~\cite{ivvsic2025diversity}.

Alternatively, simpler discrete methods can be utilized.
In one dimension, a minimal cell structure can be represented by three beads connected by springs, incorporating friction-like adhesion forces between the cell and the substrate~\cite{mai2020hydrodynamic}. 
A similar concept can be extended to two dimensions employing a network of elements connected by nonlinear springs to model how biochemical signals and intracellular forces are converted into directional motion~\cite{tarama2022mechanochemical}.
Also representations of migrating cells as three-dimensional membrane vesicles have been proposed~\cite{sadhu2021modelling}.
In this framework, the vesicle is described by a closed, triangulated surface that contains curved membrane protein complexes.
They exert protrusive forces on the membrane, representing the activity of the actin cytoskeleton, and self-organize into sheet-like lamellipodia.

Finally, a minimal continuum description of cell contour dynamics was recently also proposed, inspired by the empirical analysis of membrane displacements in terms of kymographs~\cite{schindler2021analysis}.
It relies on a stochastic component that drives the formation of membrane protrusions combined with two deterministic terms accounting for membrane retraction~\cite{schindler2024three}.

\paragraph{Cellular Potts Model}
%
The Cellular Potts Model (CPM), in contrast, involves discretizing the medium into small voxels that are assigned to specific cells or the surrounding environment, see Table~\ref{Tab1}.
Using a dynamical procedure similar to cellular automata to govern voxel updates, this method can replicate various T-cell migration modes~\cite{wortel2021local}.
These models can be extended to three dimensions, incorporating structural components such as the nucleus and lamellipodia to simulate mesenchymal cell migration on flat substrates \cite{fortuna2020compucell3d}.
They have also been successfully extended to model the collective dynamics in large ensembles of cells and tissues~\cite{thuroff2019bridging}.





\begin{table}[t]
\centering
\begin{tabular}{| p{0.16\textwidth} |  p{0.28\textwidth} | p{0.28\textwidth} | p{0.28\textwidth}  | } 
  \hline
  Feature & Phase Field & Finite Element methods & Potts models \\ 
  \hline
  Morphological Flexibility  & High 
  (large deformations and splitting) & Low  (frequent remeshing) & High 
(Can handle extreme shape changes) \\
\hline
Mathematical Complexity &
Moderate (PDE-based) &
High (Complex meshing/linear algebra) &
Low: 
(Based on Monte Carlo steps) \\
\hline 
Computational Cost &
High 
(must solve the entire grid) &
Low 
(focused on the cell domain) &
Moderate 
(cost scales with lattice resolution) \\ \hline 
Accuracy at Boundary &
Moderate (spread over a few pixels) &
High  (Explicitly follows the boundary) &
Moderate 
(Boundary depends on lattice resolution) \\ \hline 
\end{tabular}
\caption{Comparison of the performance of the main computational models for the coupling between the biochemistry and the shape of the membrane. }\label{Tab1}
\end{table}

\subsection{Modeling the center-of-mass motion of the cell}
\label{subsec3.3}

Deformations of the cell contour that are driven by intracellular cytoskeletal activity will result in a translocation of the cell's center of mass.
In most cases, center-of-mass trajectories are readily accessible through time-lapse microscopy recordings combined with subsequent image segmentation and particle tracking.
Consequently, the dynamics of the center of mass is one of the most widely used quantities to characterize cellular motility.
Typically, large ensembles of trajectories are recorded and analyzed in terms of mean squared displacements, correlation functions, and other statistical measures.
The corresponding models at the single particle level are Langevin type equations
\begin{equation}
    \frac{d\boldsymbol{v}}{dt}=\boldsymbol{F}(\boldsymbol{v})+\boldsymbol{\sigma}(\boldsymbol{v})\boldsymbol{\eta}(t) \, ,
    \label{eq:langevin}
\end{equation}
where $\boldsymbol{F}(\boldsymbol{v})$ denotes deterministic forces and $\boldsymbol{\eta}(t)$ is a 2D Gaussian white noise, with $ \langle \eta_i(t) \rangle = 0$  and $ \langle \eta_i(t) \eta_j(t') \rangle = \delta_{ij} \delta(t-t')$, where $\delta_{ij}$ is the Kronecker delta, and $\delta(t-t') $ is the Dirac delta function. 
In general, the amplitudes of the stochastic fluctuations $\boldsymbol{\sigma}(\boldsymbol{v})$ may be velocity dependent.
The simplest and most widely used form is the persistent random walk model with linear damping and additive noise
\begin{equation}
    \frac{d\boldsymbol{v}}{dt} = - \frac{1}{\tau_p}\boldsymbol{v} + \sigma \boldsymbol{\eta}(t) \, ,
\label{eq:ou}
\end{equation}
where $\tau_p$ is the persistent time.
Various more complicated functional forms of $\boldsymbol{F}(\boldsymbol{v})$ have been considered, such as quadratic velocity dependencies as well as extensions that incorporate cell-cell interactions to enable the study of collective cell migration.

A specific focus of recent research was to include external forces that represent environmental cues such as chemoattractant gradients and to use advanced methods of statistical inference to derive the underlying stochastic model from experimental data~\cite{dieterich2022anomalous}.
For a more in-depth review, see Ref.~\cite{bruckner2024learning}. 
When interacting with passive microparticles, more complex trajectory patterns may arise, such as intermittent switching between two states of locomotion~\cite{lepro2022optimal}.
Also non-Gaussian statistics can emerge as a consequence of heterogeneities in such biohybrid mixtures of motile cells and passive microparticles~\cite{grossmann2024non}.


\subsection{Modeling of collective cell motion}
\label{subsec3.4}

Having discussed individual cell migration, we now briefly turn to the interactions between multiple cells. A primary approach involves extending the single-cell phase-field model to include intercellular repulsion~\cite{alert2020physical}. In a confined domain, this repulsion is sufficient to trigger cluster formation, driven by the deformation and frustration dynamics of moving cells~\cite{moreno2022single}, or to facilitate other forms of collective behavior, such as translational collective migration~\cite{moure2021phase}.
Cells may also respond to chemotactic signals secreted by their neighbors, leading to complex emergent dynamics. These coupled internal and external biochemical processes can be effectively modeled using phase fields. In such frameworks, the external field is defined as the complement of the internal field, providing a seamless transition from individual to collective chemotactic motion~\cite{paspunurwar2024dynamic}.

In the case of {\it Dictyostelium discoideum}, also cytofission events have been successfully described in a phase-field framework.
When a small number of {\it Dictyostelium} cells are subjected to electric shocks, nanopores form in their membranes, leading to cell-cell fusion and the creation of multinucleated giant cells. These giant cells are inherently unstable and eventually undergo cell cycle-independent cytofission. Under these conditions, the phase-field method is particularly effective in modeling the complex topological changes associated with giant cell fission \cite{flemming2020cortical}.
On the other hand, to model the formation of multicellular slugs during the aggregation phase of {\it Dictyostelium}, much larger numbers of cells have to be described. For populations of this size, a discrete model is typically used, where each cell is individually represented along with explicit cell-cell adhesion terms, to simulate the dynamics of the slug~\cite{song2023differential}.

\subsection{Modeling interactions with interfaces and external boundaries}
\label{subsec3.5}

Cells interact dynamically with external objects and boundaries and may respond to their geometry.
In these scenarios, employing a phase-field model to represent the cell, the external objects, or the surrounding geometry, as for example in a microfluidic channel, is highly effective.
For example, this approach was recently used to model the oscillatory motion of cells confined between two chambers~\cite{zadeh2024inferring}, successfully replicating experimental results observed in MDA-MB-231 cells \cite{bruckner2019stochastic}.
To understand the interactions of motile cells with complex geometries~\cite{winkler2019confinement}, surface ridges~\cite{honda2021microtopographical}, or topographically patterned substrates~\cite{herr2022spontaneous}, the phase field approach was also extended to three dimensions.
Similarly, phase-field models have facilitated the study of intercellular interactions within tissues~\cite{chiang2024multiphase} and of the mechanisms by which a cell escapes from a tissue mass~\cite{melo2023ecm}.

Also minimal models of membrane vesicles driven by curved membrane protein complexes can account for complex geometrical responses, such as the directed movement along curved surfaces (curvotaxis)~\cite{sadhu2024minimal} or the interaction with external barriers~\cite{sadhukhan2025development}.



\section{Outlook on future modeling of cell crawling }
\label{sec4}




Experimental observations of cell motility are fundamentally governed by reaction rates and diffusion coefficients. Although the models discussed previously utilize parameters designed to replicate these experimental dynamics, stochastic fluctuations can significantly alter the system's behavior, often masking underlying deterministic tendencies. Recently, substantial efforts have been directed toward accounting for these stochastic disruptions, leading to the development of methods that quantify these effects \cite{abubaker2025learning} and estimate physical parameters directly from simulations \cite{pasemann2021diffusivity}.

Furthermore, cell modeling is not restricted to crawling on rigid substrates. Preliminary models of cellular swimming have been developed that consider the morphology of {\it Dictyostelium discoideum} coupled with the hydrodynamic interactions of the surrounding fluid \cite{wang2021mathematical}.

Advances in computational power and algorithmic optimization now enable the study of biological systems on multiple scales, ranging from the high-resolution dynamics of a single cell on a solid surface to the complex interactions of vast cell populations within a tissue.

In summary, the transition from experimental observation to mathematical modeling highlights the complexity of cellular motility on various scales. By integrating reaction-diffusion systems, mechanical feedback loops, and mass-conservation constraints within frameworks such as phase-field or finite element methods, we can successfully replicate behaviors ranging from individual cell polarization and to collective tissue dynamics and chemotaxis. The combination of advanced computational algorithms and multi-scale modeling continues to provide relevant information into how biochemical signals and mechanical forces converge to drive the life cycle of cells in experimental environments.

\section*{
Declaration of generative AI and AI-assisted technologies}
During the preparation of this work the authors used Gemini in order to improve
the readability of the manuscript. After using this tool/service, the authors reviewed and edited the content as needed and take full responsibility for the content of the published article.

\section*{
Declaration of Competing Interest}
The authors declare that they have no known competing
financial interests or personal relationships that could
have appeared to influence the work reported in
this paper.

\section*{Acknowledgements}
The authors acknowledge support from grant PID-2022-139215NB-I00 funded by Ministerio de Ciencia, Innovación y Universidades (MICIU/AEI/10.13039/501100011033) and by ‘ERDF: A way of making Europe’, by the European Union.
This research has been partially funded by Deutsche Forschungsgemeinschaft (DFG): Project-ID 318763901–SFB1294.

\section*{Data availability}
No data was used for the research described in
the article.





  \bibliographystyle{elsarticle-num} 
  \bibliography{COSB}





\end{document}